\documentclass[letterpaper,iop,numberedappendix]{emulateapj}
\usepackage[colorlinks=true,linkcolor=blue,anchorcolor=blue,citecolor=blue,urlcolor=blue]{hyperref}
\usepackage{color}
\usepackage{graphicx}
\usepackage{epsfig}
\usepackage{natbib}
\usepackage{rotate}
\usepackage{afterpage}
\usepackage{amsmath}
\usepackage{amssymb}
\usepackage{relsize}
\usepackage{epsfig}
\usepackage{tabularx}
\usepackage{longtable}
\usepackage{ltxtable}
\usepackage{txfonts}
\usepackage{natbib}
\newcommand{\beq}{\begin{eqnarray}}
\newcommand{\be}{\begin{equation}}
\newcommand{\ben}{\begin{enumerate}}
\newcommand{\bi}{\begin{itemize}}
\newcommand{\eeq}{\end{eqnarray}}
\newcommand{\ee}{\end{equation}}
\newcommand{\ei}{\end{itemize}}
\newcommand{\een}{\end{enumerate}}

\newcommand{\cref}{Chapter\thinspace\ref}

\newcommand{\bmp}{\begin{minipage}}
\newcommand{\emp}{\end{minipage}}

\newcommand{\ensav}[1]{\left\langle #1 \right\rangle}

\newcommand{\mr}{\mathrm}

\def\apj{ApJ}
\def\apjl{ApJL}
\def\apjs{ApJS}
\def\mnras{MNRAS}

\def\nat{Nature}

\def\physrep{Phys.~Rep.}   
\def\prd{Phys.~Rev.~D}      

\def\jcap{JCAP}

\lefthead{Krause et al.}
\righthead{The Gravitational Lensing Signal of Stacked Voids}

\begin{document}

\title{The Weight of Emptiness:\\The Gravitational Lensing Signal of Stacked Voids}

\author{Elisabeth Krause\altaffilmark{1,2}}
\author{Tzu-Ching Chang\altaffilmark{3}}
\author{Olivier Dor\'e\altaffilmark{4,1}}
\author{Keiichi Umetsu\altaffilmark{3}}
\altaffiltext{1}{Caltech, Department of Astrophysics, MC 249-17, Pasadena, CA 91125, USA}
\altaffiltext{2}{University of Pennsylvania, Department of Physics and Astronomy, Philadelphia, PA 19104, USA}
\altaffiltext{3}{IAA, Academia Sinica, P.O. Box 23-141, Taipei 10617, Taiwan}
\altaffiltext{4}{NASA Jet Propulsion Laboratory, California Institute of Technology, 4800 Oak Grove Drive, Pasadena, California, USA}

\date{\today}


\begin{abstract}
The upcoming new generation of spectroscopic 
galaxy redshift surveys will provide large 
samples of cosmic voids, the distinct, large underdense structures in the universe. 
Combining these
with future galaxy imaging surveys, we study the prospects of probing the underlying matter
distribution in and around cosmic voids via the weak gravitational 
lensing effects of stacked voids, utilizing both shear and magnification information.
The statistical precision is greatly improved by stacking together a large number of voids along
different lines of sight, even when taking into account the impact of
inherent miscentering and projection  effects.  We show that Dark
Energy Task Force (DETF) Stage IV surveys, such as the Euclid
satellite and the Large Synoptic Survey Telescope (LSST), should be
able to detect the void lensing signal  with sufficient precision from
stacking abundant medium-sized voids, thus providing direct
constraints on the matter density profile of voids independent of assumptions on galaxy bias.
\end{abstract}

\section{Introduction}
As revealed by galaxy redshift surveys, the large-scale structure of
the universe hierarchically grows into a complex network of filaments
and galaxy clusters, separated by large, nearly empty voids in
between. While the bound structures contain most of the mass in the
Universe, most of the cosmic volume is filled with voids. These voids
originate from local minima of the primordial density field and expand
faster than the average density of the Universe, clearing out their
central regions and building up filaments at their boundaries.

Just as the more studied galaxy clusters, the formation and evolution of voids probe the
extreme tails of the cosmic density distribution, and provide
insights into initial conditions and the expansion history of the Universe. For
example, the abundance of massive clusters and large voids increases
in $f(R)$ gravity models \citep{Li12}, and non-Gaussian initial
conditions change the void and cluster abundance in opposite
directions \citep{Kamionkowski09}.

Besides, as the dynamics of underdense regions is dominated by dark energy at earlier
times than the rest of the universe, the potential of voids to probe
the nature of dark energy has been noted recently. \citet{LW12} proposed an
Alcock-Paczynski test on the average shape of stacked
voids and found that it may outperform the Baryonic Acoustic
Oscillation effect as a cosmological probe for a Euclid-like
survey. In addition, the distribution of void ellipticities has been demonstrated to be a
powerful probe of the equation of state of dark energy \citep{Park07,
LW10,Bos12}. Similarly, the emptiness of voids and its evolution
over cosmic time can probe the expansion history of the universe and
modified gravity \citep[e.g.,][]{Farrar04,Nusser05,Peebles10}.

Observationally, voids are identified in the distribution of galaxies
\citep[e.g., ][, and references therein]{Sutter12}, and the matter
density profile of voids is inferred from the galaxy distribution in
and around voids. The latter requires assumptions on galaxy biasing,
which is not precisely known and subject to uncertainties in the
efficiency of galaxy formation and evolution in underdense
environments.

In this \emph{letter} we demonstrate that the matter density profile of voids
can instead be constrained directly from the average lensing signal of
stacked voids. While weak lensing by voids was previously discussed by
\citet{Amendola99}, these authors focused on the tangential shear signals by individual voids
and concluded that void radii larger than $\sim 100$\,Mpc$\,h^{-1}$ were
required to detect the effect. Compared to their analysis, stacking
allows us to achieve much higher significance in measuring the weak
lensing signals induced by more abundant, smaller voids.
large number of voids will be available from future spectroscopic
galaxy redshift surveys, such as the proposed Prime Focus Spectrograph
\citep[PFS;][]{PFS}
instrument on the Subaru telescope and the Euclid satellite;  while several future imaging
surveys, e.g., the on-going Subaru Hyper Suprime Cam (HSC), Euclid, and the LSST projects, will allow us to measure the
stacked weak lensing signals by voids. Such direct measurements of void density profiles enable cosmological
constraints without the uncertainties from galaxy biasing models, and
may additionally help in studying the environment dependence of galaxy
evolution. 

\section{Methods and Models}
\label{sec:setup}

We now forecast the sensitivity of stacked weak lensing of voids and its
ability to measure the average density profile of voids.  We consider
voids in the redshift range of $0.4<z_{\mr l}<0.6$, assuming the void
catalog is available in a survey area of $A=5000$ square degrees.  We
also assume a DETF Stage IV deep imaging survey \citep{DETF} of the
same area for  the lensing analysis; the mean background density of
galaxies is assumed to be $n_{\mr{gal}} = 12$ arcmin$^{-2}$, with a
mean redshift of $z_{\mr s}\sim 1$.  
Additionally, we consider two specific upcoming surveys: the combination of HSC and PFS, 
known as the SuMIRe survey\footnote{\texttt{http://sumire.ipmu.jp/}},
with $A = 1500$ square degrees, $n_{\mr{gal}} = 20$ arcmin$^{-2}$,  $z_{\mr s}\sim 1$,
and a Euclid/WFIRST like survey with 
$A = 15000$ square degrees, $n_{\mr{gal}} = 30$ arcmin$^{-2}$,  $z_{\mr s}\sim 1.2$.

Throughout the
paper we use the cosmological parameters derived from the {\it
  Wilkinson Microwave Anisotropy Probe} (WMAP) 7-year results
\citep{WMAP7}.   

\subsection{Void Models}
While voids are intrinsically ellipsoidal, we assume the averaged void
density profile after stacking to be spherically symmetric. In our
basic model, the void density profile and abundance follow the
simulation results of \citet{LW12}. Specifically, we use their fitting
formula 
\be 
\rho (r, R_{\mr V}) = \bar{\rho} \left( A_0 +A_3
  \left(\frac{r}{R_{\mr V}}\right)^3\right) \approx \bar{\rho}
\left(0.27 +0.61 \left(\frac{r}{R_{\mr V}}\right)^3\right)\,, 
\ee 
to describe the void density profile within $r =R_{\mr V}$, the void radius; here
$\bar{\rho}$ is the mean cosmic matter density,
and where we have ignored the scatter in the best fit parameters, 
$A_0=0.27\pm0.01$ and $A_3=0.61\pm0.03$, 
which has negigable impact on our signal-to-noise calculation,
assuming that the scatter is close to symmetric.

\begin{figure}
\includegraphics[width = 0.45\textwidth]{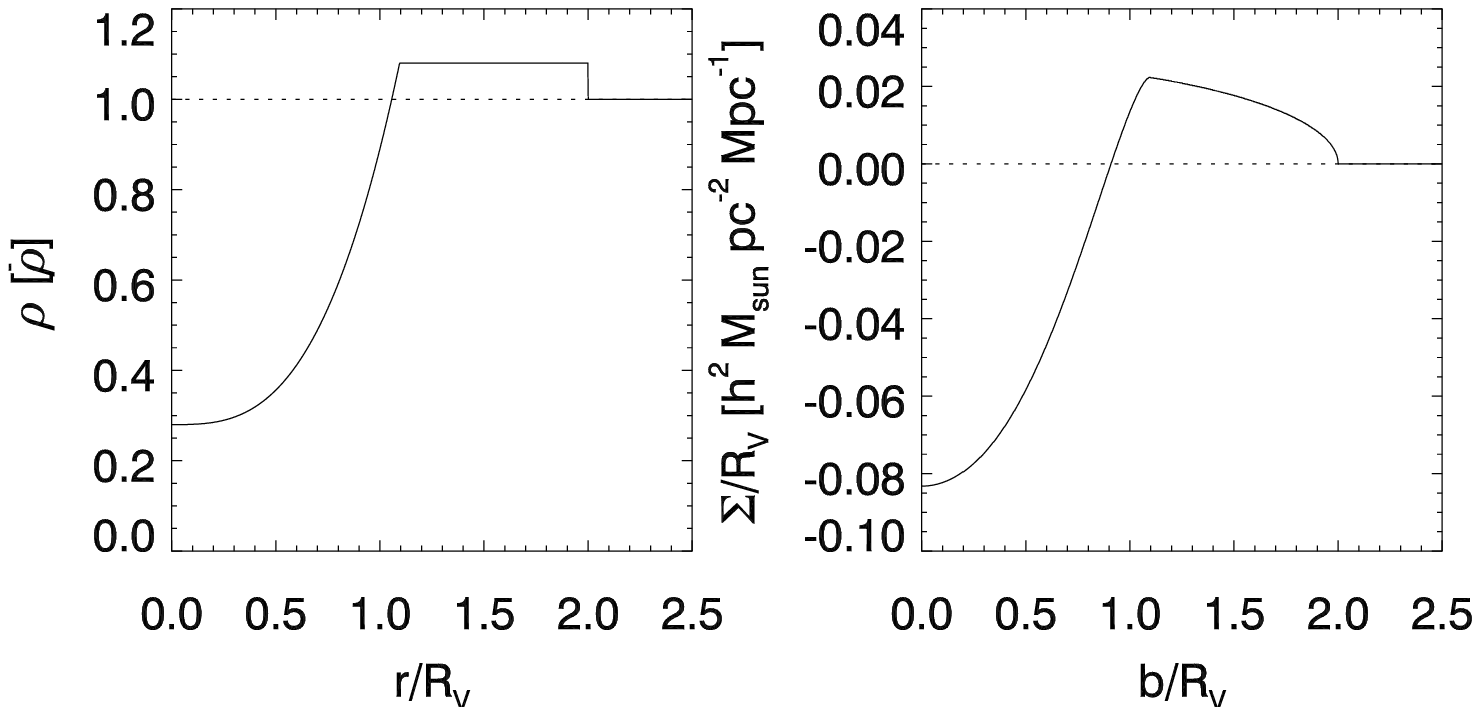}
\caption{Left: void density profile. Right: projected void surface density scaled by $R_{\mr{V}}$.}
\label{fig:profileLW}
\end{figure}

Since there are few constraints on the outer density profile of voids,
we \emph{choose} a continuation outside $R_{\mr V}$ with a constant
density wall such that profile is compensated within $2R_{\mr V}$, see
Fig.~\ref{fig:profileLW}, in order to model the extended, marginally overdense
wall structures found to surround voids in recent simulations
\citep[e.g.,][]{voidfinder, LW12} and void catalogs \citep{Sutter12}. We
explore in Section \ref{sec:discussion} a range of other models,
including voids surrounded by steep, overdense ridges, as found in
earlier simulations and void catalogs \citep[e.g.,][]{PVH12}.
The comoving number density of voids in a $1\mathrm{Mpc}h^{-1}$ wide size bin
is given by \citep{LW12}
\be
\frac{n_{\mr{void}}(R_{\mr V}\in[r,r+1\mr{Mpc}/h])}{(\mr{Mpc}/h)^{3}} = 3.5\times10^{-3}\exp\left(-0.632\frac{r}{\mr{Mpc}/h}\right)\,. 
\label{eq:nv}
\ee
\subsection{Weak Lensing Signals}
\label{sec:wl}
For a lens at redshift $z_{\mr{l}}$ the projected surface mass density $\Sigma$ is given by 
\be
\Sigma(\theta,R_{\mr V})=\int d r_z\left(\rho\left(\sqrt{(\theta D_{\mr{A}}(z_\mr{l}))^2+r_z^2},R_{\mr{V}}\right)-\bar{\rho}\right)\,,
\ee
where $D_{\mr{A}}(z)$ is the proper angular diameter distance.
In the linear regime, which is an excellent approximation for the very weak lensing caused by underdense regions, the mean tangential shear $\gamma_{\mr{t}}$ and magnification $\mu$ profile of an azimuthally symmetric mass distribution are given by
 \be
\ensav{\gamma_{\mr{t}}}(\theta)=\frac{\Delta\Sigma(\theta)}{\Sigma_{\mr{crit}}},\;\;\;\;
\mu(\theta)\approx1+2\kappa(\theta)=1+2\frac{\Sigma(\theta)}{\Sigma_{\mr{crit}}}\,,
\ee
where $\kappa$ is the lensing convergence and $\Delta\Sigma$ is the
excess surface mass density defined as
\be
\Delta\Sigma(\theta)=\bar{\Sigma}(\theta)-\Sigma(\theta)=\Sigma_{\mr{crit}}\ensav{\gamma_{\mr{t}}}(\theta)\,.
\label{eq:DS}
\ee
Here $\Sigma_{\mr{crit}}$ is the critical surface mass density weighted
by the source galaxy distribution $dn_{s}/dz_s$, 
\be
\Sigma_{\mr{crit}}=\frac{c^2}{4\pi G}\int_{z_{\mr{s,min}}}^\infty\!\!\!\!\mr{d}z_{\mr{s}}\frac{\mr{d}n_{\mr{s}}(z_{\mr{s}})}{\mr{d}z_{\mr{s}}}\frac{D_{\mr{A}}(z_{\mr{s}})}{D_{\mr{A}}(z_{\mr{l}})D_{\mr{A}}(z_{\mr{l}},z_{\mr s})}\left(\int_{z_{\mr{s,min}}}^\infty\!\!\!\!\mr{d}z_{\mr s}\frac{\mr{d}n_{\mr{s}}(z_{\mr{s}})}{\mr{d}z_{\mr s}}\right)^{-1}\,,
\ee
where we have neglected the narrow redshift distribution of the lens
 population and use the mean redshift, $z_{\mr{l}}=0.5$, in
 practice. While \citet{Bolejko12} recently pointed out that the relativistic Doppler term arrising from correlated peculiar velocities can have a significant effect on the magnification signal of source galaxies closely associated with the lensing void, this effect is negligable for our calculation due to the large separation between sources and voids, and the large number of voids used in the stacking process.

Observationally, the tangential shear is estimated as the mean
tangential ellipticity of source galaxies with respect to the void
center \citep[e.g.,][]{BS2001}.
A number of techniques have been proposed to measure the magnification
profile of galaxy groups and clusters; they broadly fall into two
categories: magnification bias measurements, which utilize the
magnification-induced variations in the apparent number density of
background galaxy populations  
\citep[e.g.][]{Hildebrandt11,Umetsu11, Ford12, Umetsu12}, and
magnification estimators based on the change in the apparent sizes and fluxes of
individual galaxies \citep[e.g.,][]{Huff11, Schmidt12}.   
Each of these estimators is affected by a variety of
systematic effects, as described in the references above; a
detailed modelling of these effects is beyond the scope of this paper, and
we will absorb this net irreducible observational noise into a simple
one-parameter model, as described below.

In order to increase the signal-to-noise ratio (S/N) of the lensing observables,
the lensing signals are measured by stacking large numbers of voids with similar sizes.  
In practice, the size measurement of an individual void in galaxy
redshift surveys
will be very noisy due to sparse sampling in the underdense region.
Hence, we follow the stacking procedure of \citet{LW12} and
\citet{Sutter12}, who stack all voids within a size bin on their
barycenters without rescaling individual voids. 
Ultimately, the uncertainties in the weak-lensing profiles are a combination of
observational errors and cosmic noise due to projection effects of
uncorrelated large-scale structure along the line of sight.
The shear (magnification) profile of voids can also be written
as the small-scale cross correlation function between void centers and shear (magnification),
and we model their covariance matrices in analogy to galaxy-galaxy lensing \citep[e.g.][]{Jeong09}.
 
The Gaussian covariance of the angular void--shear cross spectrum, $C^{\mr{V}\gamma_{\mr{t}}}(l)=C^{\mr{V}\kappa}(l)$ is given by
\beq
\nonumber\mr{Cov}\left(C^{\mr{V}\kappa}(l),C^{\mr{V}\kappa}(l')\right)&=& 
\frac{4\pi\delta^{\mathrm K}_{l,l'}}{A(2l+1)}\Bigg[
\left(C^{\mr{V}\kappa}(l)\right)^2+C^{\mathrm{VV}}\left(C^{\kappa\kappa}(l)+\frac{\sigma_\epsilon^2}{n_{\mr{gal}}}\right)\\
&&+\frac{1}{n_{\mr{V}}}
\left(C^{\kappa\kappa}(l)+\frac{\sigma_\epsilon^2}{n_{\mr{gal}}}\right)
\Bigg]\,,
\label{eq:Cvk}
\eeq
with $C^{\kappa\kappa}(l)$ the angular cosmic shear power spectrum, $\sigma^2_{\epsilon}$ the ellipticity dispersion, 
$n_{\mr{V}}$ the projected number density of voids, and
$C^{\mr{VV}}(l)$ the angular void power spectrum.
Note that the last term in Eq.~\ref{eq:Cvk} is the standard covariance estimate for stacked profiles \citep{H03}. 
In comparison, the correlation function covariance has additional contributions from the void profile ($C^{\mr{V}\kappa}(l)$) 
and from void correlations ($C^{\mr{VV}}(l)$), which increase the statistical uncertainty.
The latter includes the effective increase in shape noise caused by the overlap of projected voids.
For the purpose of this S/N estimate we assume voids to be unclustered; then $C^{\mr{V}\kappa}$ 
is given by the Fourier transform of the void convergence profile, and  $C^{\mr{VV}}$ 
is determined by its ``one-halo'' term \citep[see][for a review of the halo model]{Cooray02}.
While the void density profiles used in the lensing calculations are
compensated within $2R_{\mathrm{V}}$, we will model
projected voids as uniform disks with radius $2\theta_{\mathrm{V}}$ in
the calculation of the void clustering power spectrum
to maximize the potential degradation caused by the projected overlap of different voids, and to avoid complications with the normalization of the halo model.
Then the one-halo term of voids with angular radius $\theta_{\mr{V}}$ and angular number density $n_{\theta_{\mr{V}}}$is given by 
\be
C^{\mr{VV}}=\frac{1}{n_{\theta_{\mr{V}}}}\left(\frac{J_1(2l\theta_{\mathrm{V}})}{l\theta_{\mathrm{V}}}\right)^2\,,
\ee
with $J_n$ the $n$-th order Bessel function of the first kind. 
For an ensemble of voids with angular radii in the range $\theta_{\mr{V}}\in[\theta_{\mathrm{V,min}},\theta_{\mathrm{V,max}}]$ and radius distribution 
$dn_V(\theta_V)/d\theta_V$ it is given by
\beq
\nonumber C^{\mr{VV}}&=&\left(\frac{1}{4\pi\int_{\theta_{\mathrm{V,min}}}^{\theta_{\mathrm{V,max}}}d\theta_V\frac{d n_V (\theta_V)}{d\theta_V}\theta_V^2} \right)^2\\
    &&\times\int _{\theta_{\mathrm{V,min}}}^{\theta_{\mathrm{V,max}}}
    d\theta_V\frac{d n_V(\theta_V)}{d\theta_V}\left(4\pi\theta_V^2\right)^2\left(\frac{J_1(2l\theta_{\mathrm{V}})}{ l\theta_{\mathrm V}}\right)^2\, .
\eeq
Finally, we approximate the covariance of the angular shear profile in bins $[\theta_i-\Delta \theta/2,\theta_i+\Delta \theta/2]$ with uniform bin-width $\Delta\theta$ as
  \begin{align}
 \nonumber \mathrm{Cov}(\gamma_{\mr t}(\theta_1, \Delta\theta),\gamma_{\mr t}(\theta_2,\Delta\theta)) = \frac{1}{2\pi A}\int d l\, l J_2(l\bar{\theta}_1) J_2(l\bar{\theta}_2)\\
\nonumber 
\times \left[
  \left(C^{\mr V\kappa}(l)\right)^2 +
\left(C^{\mathrm{VV}}(l)+\frac{1}{n_{\mr{V}}}\right)C^{\kappa\kappa}(l) + C^{\mathrm{VV}}(l)\frac{\sigma_\epsilon^2}{n_{\mr{gal}}}\right] \\
+ \frac{ \delta^{\mathrm K}_{\theta_1,\theta_2} \sigma_\epsilon^2}{2 \pi A\theta_1 \Delta\theta n_V n_{\mr{gal}}}\,,
\label{eq:Cg}
\end{align}
where the Fourier transform is evaluated at the area weighted bin centers $\bar{\theta}_i$. 

Analogously, the covariance of the angular magnification profile is given by
  \begin{align}
 \nonumber \mathrm{Cov}(\mu(\theta_1, \Delta\theta),\mu(\theta_2,\Delta\theta)) = \frac{1}{2\pi A}\int d l\, l J_0(l\bar{\theta}_1) J_0(l\bar{\theta}_2)\\
\nonumber 
\times \left[
 \left(2 C^{\mr V\kappa}(l)\right)^2 + \left(C^{\mathrm{VV}}(l)+\frac{1}{n_{\mr{V}}}\right)4C^{\kappa\kappa}(l) + C^{\mathrm{VV}}(l)\frac{\sigma_\mu^2}{n_{\mr{gal}}}\right] \\
+ \frac{ \delta^{\mathrm K}_{\theta_1,\theta_2} \sigma_\mu^2}{2 \pi A\theta_1 \Delta\theta n_V n_{\mr{gal}}}\,.
\label{eq:Cm}
\end{align}
In the following discussion, we will refer to the two terms in Eqs.~(\ref{eq:Cg},\ref{eq:Cm}) proportional to $C^{\kappa\kappa}$ as large-scale structure noise, and to those proportional to $\sigma_{\epsilon/\mu}$ as observational noise. The contribution from the void lensing profile ($C^{\mr V\kappa}$) to the covariance is negligible.

Due to the high redshift of source galaxies, we assume $\sigma_\epsilon = 0.3$ per shear component.
The observational noise parameter $\sigma_\mu$ for the magnification
measurement depends on the estimator under consideration: for
magnification-bias based estimators, $\sigma_\mu = 2/(q f^{1/2}_{\mr
s})$, where $f_{\mr s}$ is the fraction of source galaxies used in the
magnification measurement, and $q \sim 1$--$1.5$ is a parameter describing
the strength of the magnification effect related to the slopes of the
galaxy luminosity and size distribution functions
\citep{Schmidt09}. For a scaling relation based estimator
\citep{Huff11}, $\sigma_\mu = 2 \sigma_\kappa/f^{1/2}_{\mr s}$, with
$\sigma_\kappa$ the scatter in the convergence estimator caused by the
scatter in the fundamental plane. $f_{\mr s}$ is the fraction of
source galaxies to which the magnification estimator can be applied,
and the factor of two propagates from convergence to magnification
error. For definiteness, we use $\sigma_\mu = 2$, but note that the
observational noise in the magnification estimate dominates over the
large-scale structure noise, so that the error bars are proportional
to $\sigma_\mu$.

\section{Results}

\begin{figure*}
\includegraphics[width = 0.9\textwidth,clip]{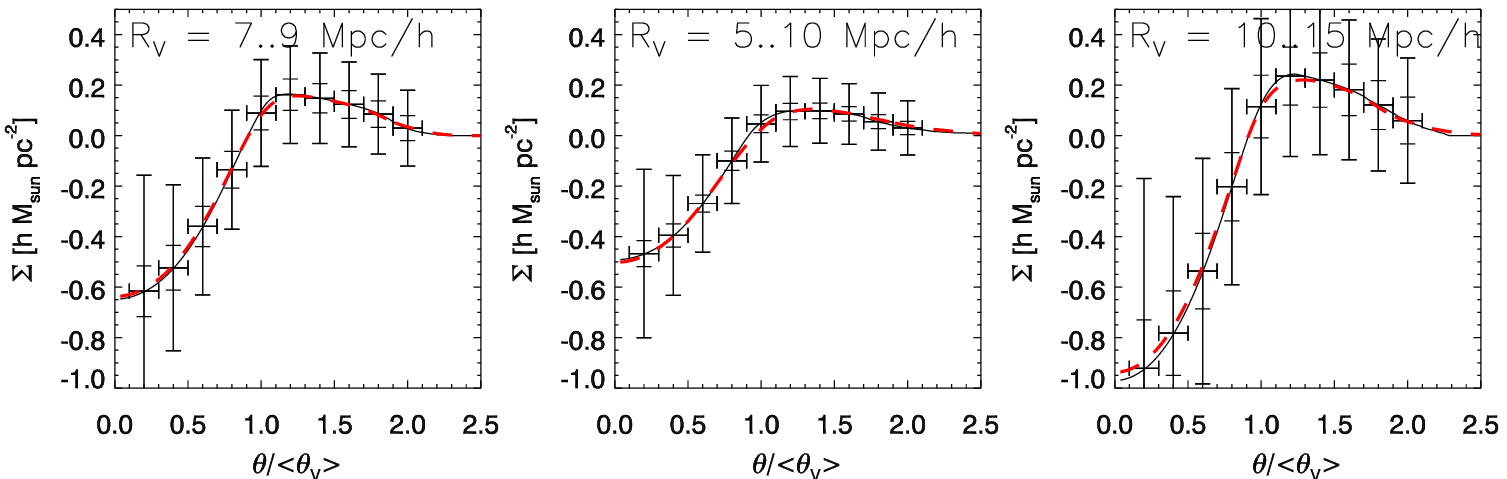}
\includegraphics[width = 0.9\textwidth,clip]{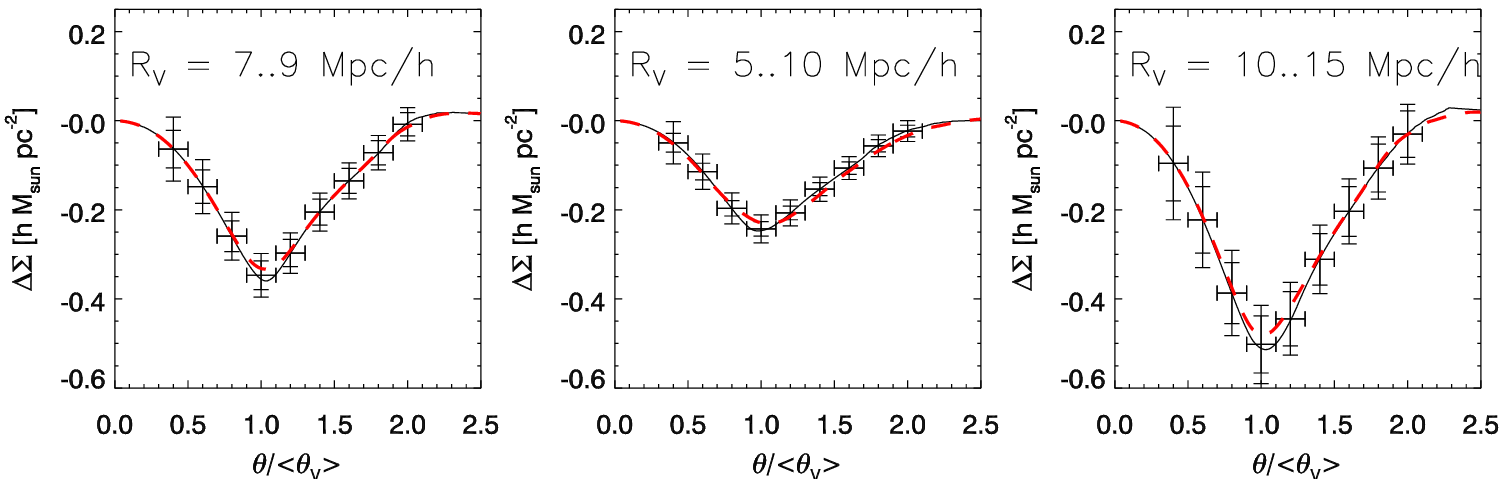}
\caption{Void lensing profiles obtained from stacking voids in different
 size bins. The top panel shows the projected surface mass density
 $\Sigma$, which is proportional to the magnification signal. The lower
 panel shows $\Delta \Sigma$ (Eq.~(\ref{eq:DS})), which is proportional
 to the mean tangential shear. The inner error bars indicate
 the cosmic noise contribution due to intervening large scale structure, 
expected for stacking voids in the redshift range $[0.4,0.6]$ 
in size bins as indicated in each plot,
and the larger error bars also include 
shape/magnification measurement noise
with survey parameters described in Sect.~\ref{sec:setup}.
The dashed lines indicate the
 degradation of the lensing signal caused by stacking voids with
 incorrectly chosen centers, assuming a two-dimensional Gaussian
 distribution of off-center positions with variance 1Mpc/$h$.} 
\label{fig:profile1}
\end{figure*}

Fig.~\ref{fig:profile1} shows the results of our calculations for the
stacked void lensing profiles in different size bins, where we have
assumed that the distribution of void sizes within each bin follows the
form of Eq.~(\ref{eq:nv}). This calculation indicates that shear and
magnification will provide strong constraints on the radial matter
density profile of small to medium sized voids, if these can be
identified from a galaxy redshift survey. 

We also include the degradation of the lensing signal by stacking voids
on incorrectly chosen centers, which is illustrated by the dashed lines
in Fig.~\ref{fig:profile1}. Specifically, we assume a two-dimensional
Gaussian distribution of off-center positions with variance
1\,Mpc\,$h^{-1}$, independent of void size. 
Note that this illustrates the effect of severe mis-centering, as it corresponds to a rms offset of 24\% of the mean void size for the smalllest void size bin ($[5,10]\,\mathrm{Mpc}/ h$).
Due to the shallow, slowly varying density of voids, off-centering has a much smaller impact on void
lensing profiles than on cluster lensing profiles.  

In Table~\ref{tab:SN} we list the radially integrated S/N for our
fiducial survey, as well as for the SuMIRe and Euclid surveys. 
Here we have included the cosmic covariance between different
radial bins,
but ignored the covariance between shear and magnification
measurements caused by large-scale structure as the magnification
measurements are strongly dominated by observational noise. 

The middle and right panels show the signal and noise expected for voids in the
radius range $[5,10]\,\mr{Mpc}/h$ and $[10,15]\,\mr{Mpc}/h$. As can be seen from
Table~\ref{tab:SN}, the integrated S/N decreases for larger voids.
In our void model the amplitude of the lensing signal is linearly proportional to void size,
while the noise increases more rapidly due to the exponentially decreasing void abundance.
Using logarithmic size bins for the stacking cannot revert this trend. Hence, for a broad range of 
steeply decreasing void abundance functions, the maximum
signal-to-noise will be obtained from the smallest idendentifiable voids.

In particular, for void sizes close to the mean spacing between galaxies,
void finding algorithms may produce incomplete void catalogs. Let $c$
denote the completeness fraction of a void catalog, such that $c = 1$ if
all existing voids are identified. Then the total S/N scales as
\be
{\mr S/N} = \left({\mr S/N}\right)_{\mr{fid}} \sqrt{c} \sqrt{\frac{A}{5000 \mr{\,sq\,dg}}}\,,
\ee
with our fiducial $\left(S/N\right)_{\mr{fid}}$ from Table~\ref{tab:SN}, which are calculated assuming $c=1$,
and $A$ the survey area.  Additionally, the S/N for magnification measurements is
approximately proportional to $\sigma_\mu$,  
\be
\left({\mr S/N}\right)_{\mu} \approx 
\left({\mr S/N}\right)_{\mu, \mr fid}\left(\frac{2}{\sigma_\mu}\right)\,
\ee
which holds as long as the error budget is dominated by observational noise ($\sigma_\mu > 0.5$).
Note that these scaling require the redshift
distribution of voids and background galaxies to stay
constant.
\begin{table}
\caption{Void number density and integrated S/N for different void
  size bins $[R_{\mr{V,min}},R_{\mr{V,max}}]$.}
\begin{center}
\begin{tabular}{l|ccc}
\hline
 & $[7,9]\,\mr{Mpc}/h$ &$[5,10]\,\mr{Mpc}/h$&$[10,15]\,\mr{Mpc}/h$\\
\hline
$n_{\mr{V}},\,[\#/\mr{sqdg}]$ &15 &72 &3\\
\hline
$\left({\mr S/N}\right)_{\mr{fid}},\,\gamma_{\mr t}$ & 12 & 18 & 10\\
$\left({\mr S/N}\right)_{\mr{fid}},\,\mu$&2 & 3 & 2\\
$\left({\mr S/N}\right)_{\mr{fid}},\,\gamma_{\mr t}+\mu$& 12 & 19 & 11\\
\hline
$\left({\mr S/N}\right)_{\mr{SuMIRe}},\,\gamma_{\mr t}$ & 7 & 10 & 5\\
$\left({\mr S/N}\right)_{\mr{SuMIRe}},\,\mu$&2 & 3 & 2\\
$\left({\mr S/N}\right)_{\mr{SuMIRe}},\,\gamma_{\mr t}+\mu$& 7 & 11 & 6\\
\hline
$\left({\mr S/N}\right)_{\mr{Euclid}},\,\gamma_{\mr t}$ & 23 & 33 & 13\\
$\left({\mr S/N}\right)_{\mr{Euclid}},\,\mu$&8 & 11 & 7\\
$\left({\mr S/N}\right)_{\mr{Euclid}},\,\gamma_{\mr t}+\mu$& 24 & 35 & 15\\
\hline
\hline
\end{tabular}
\end{center}
\label{tab:SN}
\end{table}%
\begin{figure*}
\includegraphics[width = 0.95\textwidth,clip]{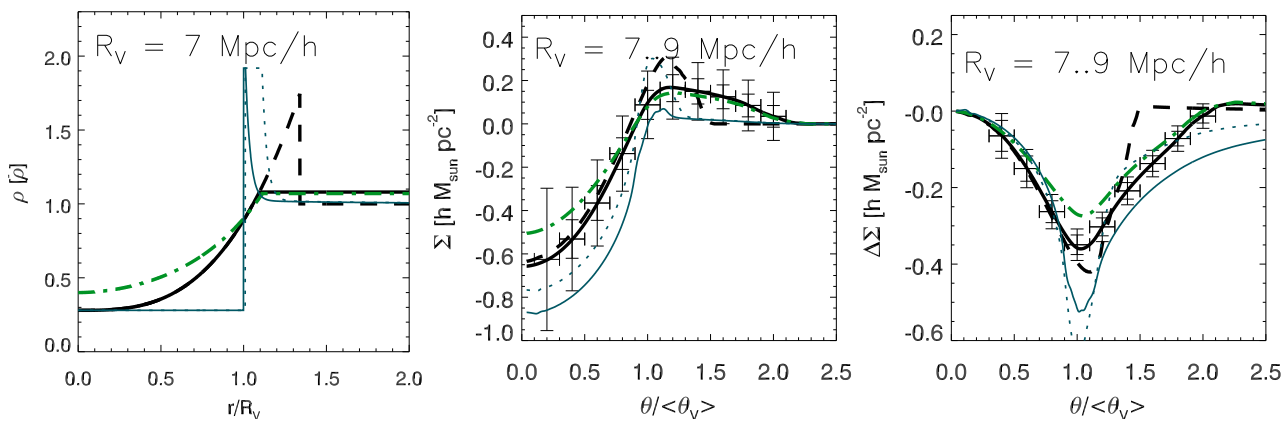}
\caption{Impact of void density profile on lensing signal. The left panel shows our fiducial void profile, which is a compensated continuation of the fit function of \citet{LW12} with an extended, low-overdensity ridge, as a thick solid line; the dashed line illustrates a compensated continuation of the same central profile with a sharp, high-density ridge; the dash-dotted line shows compensated continuation of the fit function of \citet{CSD05} with an extended, low-overdensity ridge, and the thin line shows an uncompensated void profile obtained from spherical expansion of an initial top-hat void profile in a $\Lambda$CDM universe. The profile illustrated by a dotted line was obtained by extending the ridge of the latter profile by hand. The middle and right panel show the stacked magnification and tangential shear signal of voids with different density profiles; error bars are the same as described in Fig.~\ref{fig:profile1}.}
\label{fig:profile2}
\end{figure*}

\section{Discussion}
\label{sec:discussion}

The largest theoretical uncertainty of our void model is the
distribution of matter around voids, that is, the transition from
underdense void to filament regions. In lack of theoretical models for
void density profiles we explore how well the lensing results can
differentiate between a range of toy models. Specifically, we consider
the following alternative models:
\begin{itemize}
\item[(a)] Fitting function of \citet{CSD05}, continued outside $R_{\mr
	   V}$ with an extended, low-overdensity ridge.
\item[(b)] Fitting function of \citet{LW12}, continued outside 
	   $R_{\mr V}$, truncated such that the void profile is
	   compensated.  
\item[(c)] Spherical top-hat density profile, evolved numerically in the  
	   $\Lambda$CDM cosmology \citep[c.f.][]{SW04}. 
\item[(d)] (c) with an artificially extended rigde.
\end{itemize}
Profile (a) is similar to our fiducial model but is less steep, having a
higher central density.  
Profiles (b) to (d) have sharp, high-density
ridges as illustrated in the left panel of Fig.~\ref{fig:profile2}. 
The resulting stacked lensing profiles are shown in the middle and right 
panels of Fig.~\ref{fig:profile2}. These results illustrate that magnification
and shear signals can measure the extent of wall-structures and detect
pronounced overdense ridges if these exist.
However, if individual voids have sharp overdense ridges, these ridges will be
smoothed out in the stacking process due of the
ellipsoidal shape of individual voids. Hence to better constrain the
transition between void and filament regions, in practice voids should be stacked
along their projected major axis instead. 

Additionally, the outer shear profile at $r\gtrsim R_{\mr V}$ indicates
whether the void profile is compensated 
or underdense: in the latter case the the profile falls off like a
negative point mass outside the angular extent of the void. 

Also note that profiles (b) to (d) have the same central void density as our fiducial model.
As the void lensing signal is affected by the outer void density profile at all projected radii,
measuring the emptiness of voids from lensing observations will require parametric models of the extended void profile.

\section{Conclusion}

We have demonstrated that a DETF Stage IV type of galaxy redshift
survey will allow a clear and robust detection of the stacked lensing signal of medium-sized voids with sufficient
precision, ${\mr S/N}\gtrsim 10$, and with  ${\mr S/N}\gtrsim 5\, (15)$ for the upcoming SuMIRe (Euclid) survey,
when shear and magnification are combined, and hence provide
strong direct constraints on the radial shape of the projected density
profile of voids.  The lensing measurement directly probes the mass
distribution in and around voids, and provide an unbiased view of
these underdense regions that occupy most of the cosmic
volume. It will also offer new opportunities to test gravity on
cosmic scales, but in under-dense regions, thus complementing
cluster based tests \citep[e.g.][]{Lombriser12}.

\acknowledgments
We thank Peter Schneider for critical comments, and acknowledge useful
discussions with Eric Huff, Bhuvnesh Jain, Peter Melchior and Fabian Schmidt.
This work is supported in part by the National Science Foundation under Grant No. 1066293 and the hospitality of the Aspen Center for Physics.
This research is supported in part by the National Science Council of
Taiwan grant NSC100-2112-M-001-008-MY3. KU acknowledges support from
the Academia Sinica Career Development Award.  Part of the research
described in this paper was carried out at the Jet Propulsion
Laboratory, California Institute of Technology, under a contract with
the National Aeronautics and Space Administration.

\end{document}